\documentclass[aps,pra,twocolumn,numerical,superscriptaddress]{revtex4-1}
\usepackage{graphicx}
\usepackage[unicode=true,pdfusetitle,
 bookmarks=true,bookmarksnumbered=false,bookmarksopen=false,
 breaklinks=false,pdfborder={0 0 0},pdfborderstyle={},backref=false,colorlinks=true]
 {hyperref}
\hypersetup{
 linkcolor=blue, urlcolor=blue, citecolor=blue}
\usepackage{tikz}
\usepackage{pgfplots}
\pgfplotsset{compat=1.8}
\usetikzlibrary{plotmarks}
\usetikzlibrary{external}
\tikzset{external/figure name = {fig}}
\usepackage{braket}
\usepackage{color} 
\usepackage{bm}
\usepackage{amsmath}
\usepackage{amsfonts}
\usepackage{soul}
\usepackage{algorithm} 
\usepackage{algpseudocode}

\edef\flag{1}

\begin{document}
\date{\today}
\title{Gradient-based reconstruction of molecular Hamiltonians and density matrices \\ from time-dependent quantum observables}
\author{Wucheng Zhang}
\affiliation{Department of Chemistry, University of British Columbia, Vancouver, B.C. V6T 1Z1, Canada}
\author{Ilia Tutunnikov}
\affiliation{AMOS and Department of Chemical and Biological Physics, The Weizmann Institute of Science, Rehovot, 7610001, Israel}
\author{Ilya Sh. Averbukh}
\affiliation{AMOS and Department of Chemical and Biological Physics, The Weizmann Institute of Science, Rehovot, 7610001, Israel}
\author{Roman V. Krems}
\email[Corresponding author: ]{rkrems@chem.ubc.ca}
\affiliation{Department of Chemistry, University of British Columbia, Vancouver, B.C. V6T 1Z1, Canada}
\affiliation{Stewart Blusson Quantum Matter Institute, Vancouver, B. C. V6T 1Z4, Canada}

\begin{abstract}
We consider a quantum system with a time-independent Hamiltonian parametrized by a set of unknown parameters $\alpha$. 
The system is prepared in a general quantum state by an evolution operator that depends on a set of unknown parameters $P$. 
After the preparation, the system evolves in time, and it is characterized by a time-dependent 
observable ${\cal O}(t)$.  We show that it is possible to obtain closed-form expressions 
for the gradients of the distance between ${\cal O}(t)$ and a calculated observable 
with respect to $\alpha$, $P$ and all elements of the system density matrix, whether for 
pure or mixed states. These gradients can be used in projected gradient descent to 
infer $\alpha$, $P$ and the relevant density matrix from dynamical observables. We combine
this approach with random phase wave function approximation to obtain closed-form expressions
for gradients that can be used to infer population distributions from averaged time-dependent
observables in problems with a large number of quantum states participating in dynamics. 
The approach is illustrated by determining the temperature of molecular gas (initially, in thermal 
equilibrium at room temperature) from the laser-induced time-dependent molecular alignment.
\end{abstract}
\maketitle

\section{Introduction}
Imaging molecular wave packets using quantum measurements represents a major challenge,
since molecules have complex energy 
levels structure, requiring large basis sets to describe dynamics. 
While several approaches, whether equivalent to quantum state tomography
\cite{qst-molecules,mandelshtam1997harmonic,avisar2011complete} or least-squares fitting \cite{Hasegawa2008qsr,
He2019Direct, karamatskos2019molecular,Ueno2021QS}, have been demonstrated, they either require sophisticated 
measurements of angle-resolved observables \cite{He2019Direct,karamatskos2019molecular,Ueno2021QS}, 
probe a restricted range of molecular states participating in dynamics \cite{Hasegawa2008qsr} or 
can only accommodate a single observable \cite{qst-molecules,mandelshtam1997harmonic,avisar2011complete}. 
A molecular imaging experiment is further complicated by experimental uncertainties in parameters
of the fields used for initial state preparation and/or for probing. It is vital to 
develop approaches for obtaining simultaneously the 
preparation/probing field parameters, the Hamiltonian parameters and quantum states that 
correspond precisely to experimental observables. This problem can, in principle, be 
addressed with a combination of quantum state and quantum process tomography (QT) aiming
to determine the density matrix and the transformations it undergoes. However, extensions 
of QT approaches to molecular dynamics are challenging. 
Filter-diagonalization techniques have been applied to extract the spectral function from time-dependent 
signals \cite{mandelshtam1997harmonic, mand1, mand2}.
An alternative approach is based on optimization of feedback loops widely employed in optimal control applications
\cite{He2019Direct, Judson1992, Peirce1988}.  Optimal control is generally concerned 
with the problem of tuning field parameters to achieve results 
equivalent to those of quantum inverse problems \cite{ip,le-roy,ip-2,rkr}. The convergence
of feedback loops can be accelerated with machine learning algorithms such as 
Bayesian optimization, which has been exploited to obtain molecule--molecule interaction
potentials in quantum reaction dynamics \cite{BO} and dipole polarizability tensors
from molecular orientation measurements \cite{hancock-paper}.
However, an extension of these approaches to quantum state reconstruction requires
the feedback parameters to describe the entire density matrix. In this case, the number
of feedback parameters grows exponentially with the number of molecular states.  

In the present work, we demonstrate an efficient and scalable approach to obtain the 
density matrix and the parameters of Hamiltonian and state-preparation fields from averaged time-dependent
observables, such as molecular alignment or a combination of alignment and orientation. 
The problem is formulated as estimation of a target vector, representing the observable at
discretized time instances, by vectors parametrized by density matrix elements. We show that the gradient
of the norm of the difference between the target observables and estimators with respect to the density 
matrix elements and Hamiltonian parameters can be evaluated analytically. Since density matrix evolution
preserves normalization and respects selection rules, we use the gradients
thus obtained in a projected gradient descent calculation, where each gradient-driven update 
is projected to the nearest point in the constrained domain. While the formalism introduced 
in the next section is general, our numerical examples treat the rotatinal motion of 
linear molecules (under rigid rotor approximation) excited by non-resonant laser pulses.
We demonstrate the approach by reconstructing
the field parameters and the density matrix of a molecular gas from noisy alignment/orientation
signals. We show that this technique allows reconstructing the density matrix, either 
before or after the laser excitation. 
For molecules at thermal equilibrium, we show that this approach can 
be used to determine the temperature by measuring molecular alignment as a 
function of time.

The present work assumes that measured time-dependent observables contain sufficient information to 
reconstruct the entire density matrix. Care must be taken to ensure this. To illustrate this point, 
we consider molecules excited by two delayed cross-polarized laser pulses. We find that one requires 
three different observables to reconstruct the resulting density matrix. Extensions of the present 
approach to include vibrational and electronic degrees of freedom may require more observables.  
Whether there exist molecular density matrices that cannot -- in principle -- be uniquely reconstructed from 
a set of time-dependent observables remains an open question that we leave for future work.

\section{Theory}
\subsection{Gradient Evaluation}
We consider a quantum system with the density operator $\hat \rho (t)$ and the 
time-independent Hamiltonian $\hat{H}({\alpha})$ yielding the propagator $\hat{U}(t)$ 
so that $\hat \rho(t) = \hat{U}(t) \hat{\rho}(t=0) \hat{U}^\dagger(t)$. Here, $\alpha$ 
is a set of parameters determining the eigenspectrum of $\hat{H}$. 
%The observable is given by  ${\cal O}(t) = {\rm Tr}[\hat \rho(t) \hat{{\cal O}}]$. 
Before interaction with an external field, the system is in an unknown mixture of quantum states 
\begin{eqnarray}
   \hat \rho_{\rm ini} =\sum_{j} p_j \ket{\chi_j}\bra{\chi_j}, 
\end{eqnarray}
where $\ket{\chi_j}$ is either an eigenstate of $\hat H$ or a coherent superposition of eigenstates of $\hat H$. 
The interaction with the laser field is assumed to be in the impulsive limit, yielding
\begin{eqnarray}\label{eq:rho(t=0)}
\begin{aligned}
   \hat \rho(t=0) &=  \hat{V}(P) \hat \rho_{\rm ini} \hat{V}^\dagger(P) \\
                  &= \sum_{j} p_j \ket{\psi_j}\bra{\psi_j} = \sum_{j} p_j \hat \rho_j,
   \end{aligned}
\end{eqnarray}%
where $\hat{V}(P)$ is the interaction-induced transformation, $P$ is a set of unknown interaction 
parameters, $| \psi_j \rangle = \hat V(P) | \chi_j \rangle$, and $\hat \rho_j = \ket{\psi_j}\bra{\psi_j}$. 
The goal is to 
infer $\alpha$, $p_j$, $P$ and $\hat \rho_j$ given an observable or a set of observables. 
Once $p_j$, $P$ and $\hat \rho_j$ are inferred, $\hat \rho_{\rm ini}$ is fully determined. 

We write 
\begin{eqnarray}
\hat \rho_j = % = \ket{\psi_j}\bra{\psi_j} =
\sum_n \sum_m c_{n,j}  c_{m,j}^\ast  \ket{\phi_n}\bra{ \phi_m},
\end{eqnarray} 
where $\ket{\phi_n}$ are the eigenstates of $\hat{H}$, and aim to determine 
the complex valued coefficients $c_{n,j}~\forall~n$ and $j$. For a given $j$, we arrange the coefficients $c_{n,j}$ in a column vector $\bm \psi_j$ and introduce $\bm a = {\rm Re}(\bm \psi_j)$ and 
$\bm b = {\rm Im}(\bm \psi_j)$, which was also used in \cite{palao2003optimal}.

The time dependence of an observable is given by
\begin{eqnarray}
    {\cal O}(t) = \sum_j p_j {\rm Tr} \left [ \hat \rho_j(t)  \hat{{\cal O}} \right ] 
    = \sum_j p_j {\cal O}_j(t).
\end{eqnarray} 
We discretize time $t$ into $K$ values and aim to minimize 
\begin{eqnarray}
    E = \sum_{c=1}^K |{\cal O}(t_c) -  {\cal O}_{\rm ref}(t_c)  |^2,
\label{eq:objective-E}
\end{eqnarray}
where ${\cal O}(t_c)$ is the observable at time $t_c$ calculated for a trial set of 
parameters and ${\cal O}_{\rm ref}(t_c)$ is an experimental measurement of ${\cal O}$ at $t=t_c$.
For the numerical examples in this work, we generate 
${\cal O}_{\rm ref}(t_c)$ by rigorous quantum calculations with known parameters. 
These parameters are then inferred using the proposed method, relying on no other information than the reference signal.   
 
%numerically generated reference signals that simulate experimental measurements.

The gradients of $E$ with respect to $p_j$, 
${\bm a}$, $\bm b$ and $P$ can be written as follows: 
\begin{eqnarray}
    \frac{\partial E}{\partial p_{j}} = 
    \sum_{c=1}^K 2 \left [ {\cal O}(t_c)-{\cal O}_{\rm ref}(t_c)\right ]{\cal O}_j(t_c),
\label{population-derivatives}
\end{eqnarray}
\begin{eqnarray}
     \frac{\partial E}{\partial {\bm a}} =  
    \sum_{c=1}^K 2\left [ {\cal O}(t_c)-{\cal O}_{\rm ref}(t_c)\right ]
    \sum_{j}p_{j} \nabla_{\bm a} {\cal O}_j, 
\label{state}
\end{eqnarray}
\begin{eqnarray}
   \frac{\partial E}{\partial {\bm b}} =  
    \sum_{c=1}^K 2\left [ {\cal O}(t_c)-{\cal O}_{\rm ref}(t_c)\right ]
    \sum_{j}p_{j} \nabla_{\bm b} {\cal O}_j, 
\label{state}
\end{eqnarray}
and
\begin{eqnarray}
\frac{\partial E}{\partial P} = \frac{\partial E}{\partial \boldsymbol{a}}
    \mathrm{Re}\left(\frac{\partial {{\bm \psi}_j}}{\partial P}\right)+\frac{\partial E}{\partial \boldsymbol{b}}
    \mathrm{Im}\left(\frac{\partial {{\bm \psi}_j}}{\partial P}\right),
\label{P-field}
\end{eqnarray}
where, for simplicity, we assume that $P$ is a single parameter characterizing
the strength of the excitation. Generalization to the case when 
$P$ is a set containing more than one parameter is straightforward.
To evaluate $\partial {\cal O}_j(t_c)/\partial {\bm a}$ and $\partial {\cal O}_j(t_c)/\partial {\bm b}$, 
we apply the results of Appendix to 
${\cal O}_j = \braket{\psi_j | \hat{U}^\dagger \hat{{\cal O}} \hat{U} | \psi_j}$ and arrive at: 
\begin{eqnarray}  
    \label{C-a}
    \nabla_{\bm a} {\cal O}_j &=& \bm{U}^{\dagger} \bm{{\cal O}} \bm{U}{\bm \psi_j}
                         + \left(\bm{U}^{\dagger} \bm{{\cal O}} \bm{U}\right)^{T}{\bm \psi_j}^{*}, \\
    \nabla_{\bm b}  {\cal O}_j &=& i\left(\bm{U}^{\dagger} \bm{{\cal O}} \bm{U}\right)^{T}{\bm \psi_j}^{*}
                          - i\bm{U}^{\dagger} \bm{{\cal O}} \bm{U}{\bm \psi_j},
\label{C-b}
\end{eqnarray}
where all operators are represented by matrices in the basis $|\phi_n \rangle$. 
%
%where $a$ and $b$ are vectors containing $a_{n,j}$ and $b_{n,j}$ from $c_{n,j}=a_{n,j}+ib_{n,j}$.

We thus obtain analytic gradients with respect to the probabilities of the mixed states $p_j$, 
the parameters of the state preparation fields $P$ and the real and imaginary parts of the 
expansion coefficients $c_{n,j}$. To ensure proper solutions, these gradients are used in constrained optimization that projects 
each gradient-informed update to the nearest point in the domain that respects: (i) the normalization of $p_j$ and 
$c_{n,j}$; (ii) selection rules; (iii) conservation of corresponding quantum numbers.

\subsection{Projected Gradient Descent}

The gradients derived in the previous section can be used in gradient descent optimization. However, because the variables of the optimization are physical parameters, it is necessary to constrain the optimization to ensure that the parameters satisfy physical conditions. 
For example, the coefficients $c_{n,j}$ or the probabilities $p_j$ must always remain properly normalized. This can be achieved by means of the projected gradient descent method.

We consider a parameter space $\mathbb{R}^n$ with parameters constrained in the domain $S \subset \mathbb{R}^n$. 
For example, in the case of probabilities, $S$ is the hyperplane $\sum_j p_j=1$.
The projected gradient descent method enables the optimization of $f(\boldsymbol{x})$ under the constrain $\boldsymbol{x} \in S$ \cite{beck2014introduction}. The algorithm can be summarized  as follows:
\begin{enumerate}
   \item Start randomly at $\boldsymbol{x}_0$. 
   \item  For $k \in \mathbb{Z} \cap [0, \infty )$ and until convergence is reached: 
   \begin{enumerate}
   \item Evaluate the gradient $\nabla f(\boldsymbol{x}_k)$.
   \item Update $\tilde{\boldsymbol{x}}_{k+1}=\boldsymbol{x}_k-\alpha_k \nabla f(\boldsymbol{x}_k)$ with step size $\alpha_k$.
   \item Project $\tilde{\boldsymbol{x}}_{k+1}$ to the constrained domain by finding $\boldsymbol{x}_{k+1}=\mathrm{argmin}_{\boldsymbol{x} \in S} |\boldsymbol{x}-\tilde{\boldsymbol{x}}_{k+1}|$. 
   \end{enumerate}
\end{enumerate}

%\begin{algorithm}
%\begin{algorithmic}[1]
%	\State Start randomly at $\boldsymbol{x}_0$. 
%		\For {$k \in \mathbb{Z} \cap [0, \infty )$ and until convergence is reached}
%		\State Evaluate the gradient $\nabla f(\boldsymbol{x}_k)$.
%		\State Update $\tilde{\boldsymbol{x}}_{k+1}=\boldsymbol{x}_k-\alpha_k \nabla f(\boldsymbol{x}_k)$ with step size $\alpha_k$.
%		\State Project $\tilde{\boldsymbol{x}}_{k+1}$ to the constrained domain by minimization $\boldsymbol{x}_{k+1}=\min_{\boldsymbol{x } \in S} \|\boldsymbol{x}-\tilde{\boldsymbol{x}}_{k+1}\|$
%		\EndFor
%\end{algorithmic}
%\end{algorithm}
%For example, the projection step involving $c_{n,j}$'s corresponds to the renormalization of $c_{n,j}$ vector, and the projection step involving $p_j$'s corresponds to the projection of $\nabla E$ onto the hyperplane $\sum_j p_j=1$.

\subsection{Random Phase Wave Function}
To construct the full quantum state at time $t$, the time propagators should be applied to 
each of the Hamiltonian eigenstates participating in dynamics. However, at high temperature, 
the number of populated energy states becomes very large, making this approach
computationally expensive. For such problems, we combine the gradients derived above with the 
random phase wave function (RPWF) approach \cite{gelman2003simulating,nest2007quantum,RPWF}. For a system with a large number of $\ket{\chi_n}$
and associated population probabilities, one can introduce a RPWF:
\begin{eqnarray}
    \ket{\alpha_k} = \sum_{n} e^{- i \alpha^n_k} \ket{\chi_n} \sqrt{p_n},
\end{eqnarray}
where the phase $\alpha^n_k$ is randomly sampled from an interval $[0, 2 \pi]$. 
The state preparation field is then assumed to apply to $\ket{\alpha_k}$ to 
produce the density operator $\hat \rho_k$ and the observable 
${\cal O}_k(t) = {\rm Tr} \left [\hat \rho_k(t) \hat{{\cal O}} \right]$. 
The average 
% over $N$ samples of $k$ 
%
\begin{eqnarray}
    \braket{\hat{{\cal O}}}_{\rm RPWF}(t) = \frac{1}{N} \sum_{k=1}^N  {\cal O}_k(t)
\end{eqnarray}
is known to converge to the exact result in the limit of large $N$ \cite{RPWF}. 
By linearity,
\begin{align}
  \braket{\hat{{\cal O}}}_{\text{RPWF}}(t) = 
%   & 
%           \frac{1}{N} \sum_{k=1}^N \braket{ p|\hat{\Theta}^\dagger_k \hat{V}^\dagger 
%           \hat{U}^\dagger(t) \hat{{\cal O}} \hat{U}(t) \hat{V}\hat{\Theta}_k|p} \nonumber\\ 
%           = 
           & \braket{ p|\hat{M}|p},
\end{align}
where ${\bm M}=N^{-1}\sum_{k=1}^N  {\bm \Theta}^\dagger_k {\bm \Pi}^\dagger {\bm V}^\dagger {\bm U}^\dagger(t) {{\bm O}}
\bm {U}(t) \bm {V}\bm \Pi\bm{\Theta}_k $, with $\bm {\Theta}_k$ being a diagonal matrix of $\exp(-i\alpha_k^{n})$, the elements of $\bm \Pi$ given by $\langle \phi_m | \chi_j \rangle$,
and $\ket{p} = \sum_j \sqrt{p_{j}}\ket{\chi_j}$. Thus the gradient can be evaluated as
\begin{flalign}
    \!\frac{\partial E}{\partial \bm{p}} 
    \! =\!& \sum_{c=1}^K 2[\braket{\hat{{\cal O}}}_{\text{RPWF}}(t_c)
    \!-\!{\cal O}_{\rm ref}(t_c)](\bm{M}\!+\!\bm{M}^\top)\!~\bm{p}, \label{rpwf}
\end{flalign}
where $\bm p$ is a column vector of $\sqrt{p_j}$.

\section{Results}
We now illustrate the proposed approach by considering several problems of increasing 
complexity. We consider an ensemble of homonuclear rigid diatomic molecules with 
$\hat{H}=\hat{J}^2/(2I)$, where $\hat{J}$ is the angular momentum operator, and $I$ is 
the moment of inertia. The spectrum (in atomic units) and eigenstates of $\hat{H}$ are $J(J+1)/(2I)$ 
and $\ket{J,M}$, where $M$ is the eigenvalue of $\hat{J}_Z$, the $Z$ component of 
$\hat{{J}}$ \cite{Zare1991,Krems2018}. We use spherical harmonics $Y_{JM}(\theta,\varphi)$ 
with $\theta$ and $\varphi$ defined as in Ref. \cite{Zare1991} to represent $|J,M\rangle$. 

\subsection{Pure State: Excitation by a Single Linearly Polarized Laser Pulse} \label{pl}
First, we consider molecules at zero initial temperature in $\ket{J,M}=\ket{0,0}$ excited by a non-resonant femtosecond 
pulse linearly polarized along the $Z$ axis. The pulse creates coherent superpositions of a large number of 
$\ket{J,M=0}$ states and induces molecular alignment \cite{Stapelfeldt2003,Fleischer2012,Krems2018,Koch2019}. 
In the impulsive approximation, 
$ V(P) = \exp[i P \cos^2(\theta)]$, where $P$ is the interaction strength parameter \cite{Fleischer2009}. 
The matrix representation of $V(P)$ in the basis $|JM\rangle$ can be evaluated using the expansion of  $\exp[i P \cos^2(\theta)]$ in spherical harmonics \citep{Leibscher2004} and 
applying the Wigner-Eckart theorem \cite{Zare1991}. 
We set $P = 8.278$. 

\begin{figure}[ht] 
    \if\flag1\includegraphics{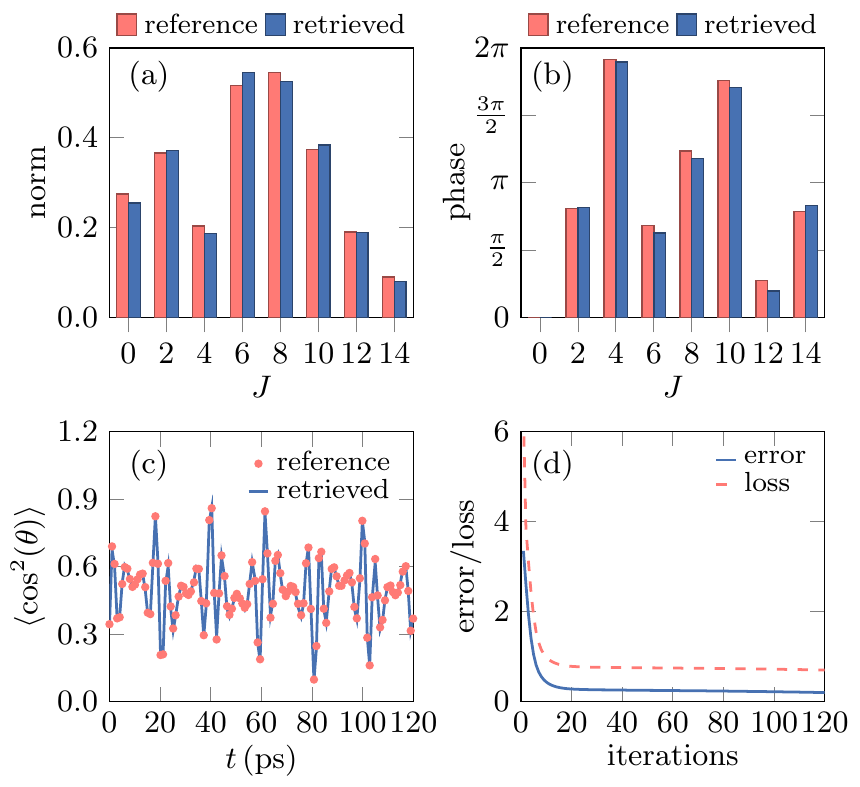}\else\include{fig1}\fi
    \vspace{-3mm}
    \caption{Comparison of reference (red) and reconstructed (blue) values of the 
    norm (a) and phase (b) of $C_{J}$. The reconstruction is based on the 
    $\braket{\cos^2(\theta)}(t)$ signal, including $3 \%$ Gaussian noise, as 
    shown in (c). Panel (d) illustrates the convergence of the target 
    vector estimation error and the loss function, i.e. error with L2 regularization.}
\label{pure-state}
\end{figure}

We compute a reference time-dependent degree of alignment quantified by $\braket{\cos^2(\theta)}(t)$ 
and modulate it by adding $3\%$ Gaussian noise to simulate an experimental signal 
${\cal O}_{\rm ref}(t)$. This reference signal is then used to reconstruct the real and 
imaginary parts of the probability amplitudes $C_J$ for each $\ket{J,M=0}$ state in the wave 
packet. To account  for noise, we regularize the objective function 
in Eq. \eqref{eq:objective-E} by adding the L2 norm $|\boldsymbol{\psi}|$, as is commonly done in ridge regression in machine learning \cite{ml-book}.
For this problem, we only require the gradients given by Eqs. (\ref{C-a}) and (\ref{C-b}). Figure 
\ref{pure-state} illustrates that projected gradient descent steered by these 
derivatives reconstructs both the norm and the phase of each wave packet component 
with as few as 10 iterations. 

\subsection{Thermal Ensembles}
\begin{figure}[ht] 
    \if\flag1\includegraphics{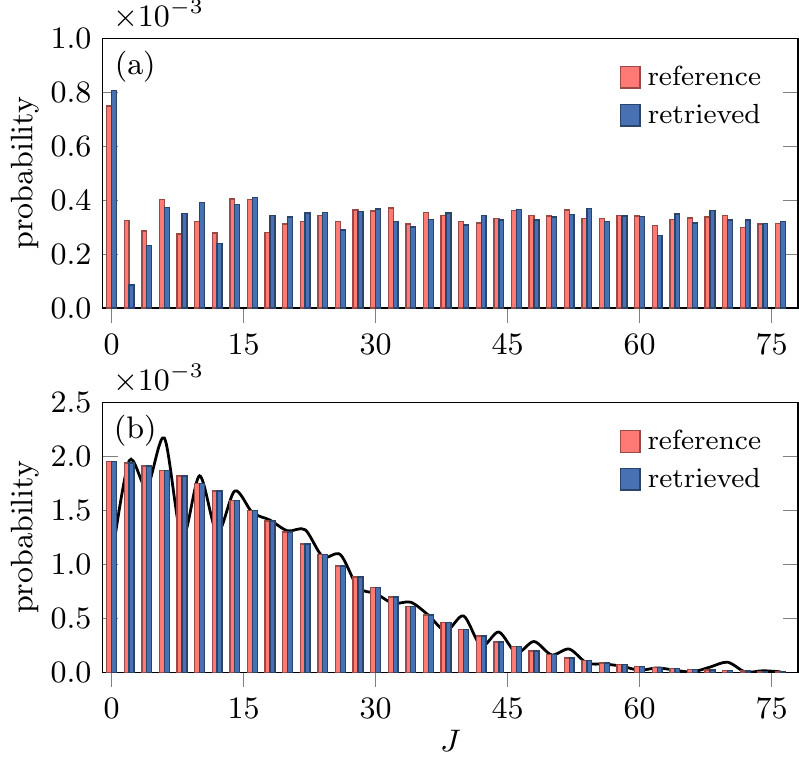}\else\include{fig2}\fi
    \vspace{-3mm}
	\caption{Comparison of reference (red) and reconstructed (blue) 
	probabilities for: (a) -- a mixture consisting of 3003 rotational states, $|JM\rangle$ with even $J \leq 76$ with
	random populations drawn from a uniform distribution; and (b) -- a thermal ensemble of molecules with 
	$T=300\,\mathrm{K}$. 
	The reconstruction is based on $\braket{\cos^2(\theta)}(t)$ 
	simulated using  the RPWF approach. The solid curve in (b) is obtained 
	without temperature gradients or any information on the relative population probabilities.
	}
\label{room-temperature}
\end{figure}

We now consider ensembles of molecules 
initially in mixed states including a large number of molecular rotational states. We consider two cases: 
(i) an ensemble of 3003 states $\ket{J,M}$ with even $J \leq 76$, where the initial populations are 
randomly drawn from a uniform distribution; and (ii) a thermal ensemble of molecules at room temperature. The reference 
signals for these ensembles are computed using the exact time propagation of each molecular state 
(up to $J=76$) participating in dynamics. The inferred probabilities are obtained using Eq. (\ref{rpwf}) based on RPWF with $N=30$, which leads to convergent results and is sufficient to recover the probability distribution.
%It has been demonstrated in \cite{RPWF} that the approximation error, $\varepsilon$ scales with the temperature $T$ and $N$ as $\varepsilon \sim 1 / \sqrt{N T^{3 / 2}}$. 
Figure \ref{room-temperature} illustrates 
the efficiency of our approach applied to two different initial population distributions.
We further illustrate the generality of our approach by the solid line in Fig. \ref{room-temperature}(b) 
showing the results for a thermal ensemble obtained without any information about relative population 
probabilities.

\begin{figure}[ht] 
    \if\flag1\includegraphics{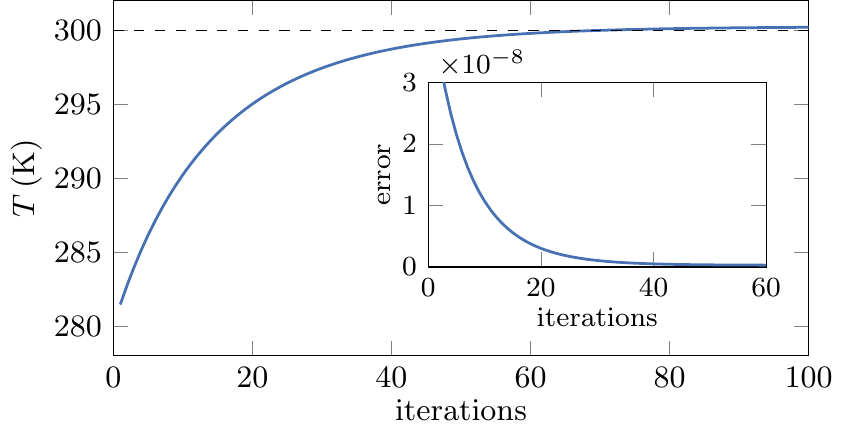}\else\include{fig3}\fi
    \vspace{-3mm}
    \caption{Convergence of  the inferred temperature and the target 
    vector estimation error (inset) for the $\braket{\cos^2(\theta)}(t)$
    signal used for Fig. \ref{room-temperature}.}
\label{room-T}
\end{figure}

For a thermal ensemble, the population probability gradients [see Eq. 
\eqref{rpwf}] can be extended by the chain rule to obtain the temperature gradient 
\begin{equation}
    \frac{dE}{dT} = \frac{dE}{d \bm p}\frac{d \bm p}{dT},
\end{equation}
yielding a formalism for determining the temperature directly from an observed laser-induced 
alignment signal. The bars in Figure  \ref{room-temperature}(b) illustrate the probabilities inferred
for a Boltzmann distribution. The difference between the solid curve and the bars shows the improvement
of the inference resulting from the inclusion of the temperature gradients. 
The convergence of inferred
temperature for the thermal ensemble in Fig. \ref{room-temperature}(b) 
is illustrated in Fig. \ref{room-T}. The slower convergence of $T$ compared to the error is due to the exponential relation $p_j \sim \exp[-E/(kT)]$ where the small change in error corresponds to the large change in $T$ near the minimum. 
A related problem of inferring rotational temperature from an alignment 
signal was recently considered experimentally in Ref. \cite{Oppermann2012}. 
The rotational temperature of molecules after impulsive excitation was obtained by least-squares fitting, limiting the application to cold gases ($< 58$ K), with accuracy of $\approx 2-4\,\mathrm{K}$, likely limited by experimental noise. 
The present approach applies to 
a wider range of $T$ and allows for control of inference accuracy through the proper choice of L2 regularization, the number of iterations and $N$. 
\subsection{Pure State Excited by Two \\Cross-Polarized Laser Pulses}
\begin{figure}[ht] 
    \if\flag1\includegraphics{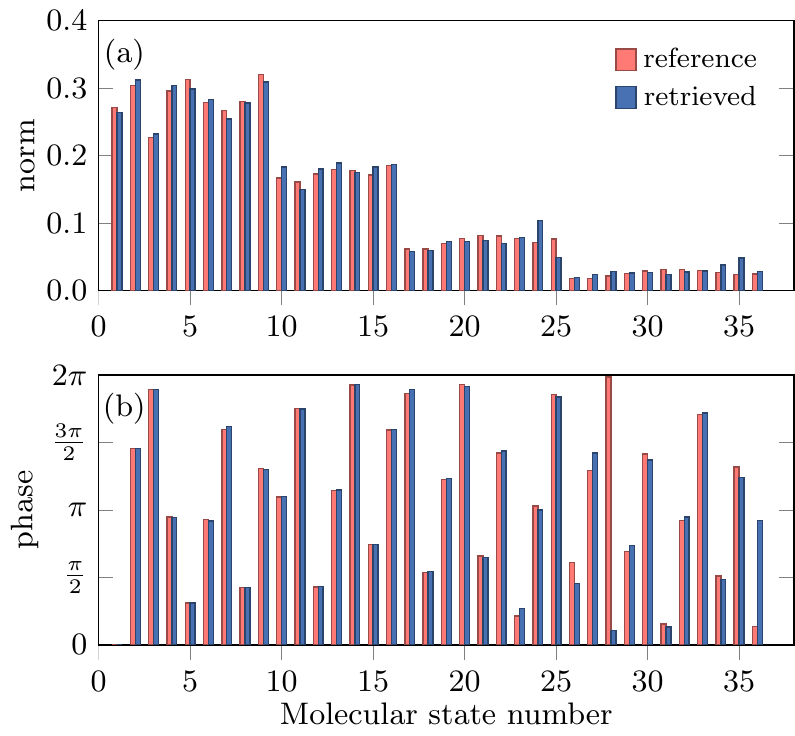}\else\include{fig4}\fi
    \vspace{-3mm}
	\caption{Comparison of reference (red) and reconstructed (blue) values
	of the norm (a) and phase (b) of $C_{JM}$ for the rotational wave packet 
	produced by excitation of molecules initially in $\ket{J,M}=\ket{0,0}$
 	with two cross-polarized pulses. The molecular states are arranged in order of increasing $J$ and, for a given $J$, in order of increasing $M$. The wave-packet reconstruction
	is based on a combination of three signals: $\braket{ \cos^2(\theta)}$, 
	$\braket{\cos^2(\phi)}$ and $\braket{\sin^2(\theta)\sin(2\phi)}$ 
	used simultaneously. Notice that the phase difference for molecular state 28 appears to be large as a consequence of constraining the phase to the interval $[0, 2\pi)$.}
\label{mixed-signal}
\end{figure}

We now consider a molecular ensemble excited by two delayed 
cross-polarized laser pulses: the first polarized along the 
$X$ axis and the second in the $XY$ plane at $45^\circ$ to the $X$ axis. Such an excitation induces molecular unidirectional
rotation \cite{Fleischer2009,Lin2015,Mizuse2015}, i.e. creates a wave 
packet with coherences between states with different $J$ and $M$. 
This significantly enlarges the corresponding density matrices and makes 
the inverse problem non-unique: a single alignment signal, $\braket{\cos^2(\theta)}$, 
is not sufficient to determine the probability amplitudes $C_{JM}$ for all $\ket{JM}$. 
The non-uniqueness is resolved by including multiple reference signals 
into the objective function in Eq. \eqref{eq:objective-E}. We find that all $C_{JM}$ can be reconstructed
if the reference signal includes simultaneously three observables: 
$\braket{ \cos^2(\theta)}$, $\braket{\cos^2(\phi)}$ and 
$\braket{\sin^2(\theta)\sin(2\phi)}$. All of these observables can be 
probed either optically or using Coulomb explosion-based methods
\cite{Renard2003,Faucher2011,Damari2016,Stapelfeldt2003,Koch2019,karamatskos2019molecular}.
Recently developed experimental techniques, such as the cold target 
recoil ion momentum spectroscopy (COLTRIMS) \cite{Dorner2000,Lin2015,Xu2020}, 
allow one to probe these observables simultaneously. 
Figure \ref{mixed-signal} illustrates the norm and phase of each of the 
coefficients $C_{JM}$ inferred from a combination of these observables. 
The molecular state number refers to $\ket{JM}$ states ordered as
$\ket{0,0},\,\ket{1,-1},\,\ket{1,0},\,\ket{1,1},\dots$
\vspace{-6mm}
\subsection{Simultaneous Reconstruction}
Finally, we consider a problem of simultaneous reconstruction of the 
Hamiltonian parameters ($\alpha$ and $P$) and the populations of a mixed state
from a time-dependent alignment signal following a single-pulse excitation. 
We treat the molecular moment of inertia $I$ and the preparation field parameter 
$P$ in a linearly polarized (along $Z$ axis) laser pulse as unknown variables
and supplement the gradients to include the derivatives with respect to $P$ and $I$. 
We write the derivatives in Eq. (\ref{P-field}) as
\begin{eqnarray}
    \frac{\partial \bm{\psi_{j}}}{\partial P} = \frac{\partial \bm{V}(P) }{\partial P} \bm{\chi_j}, 
%\\    i \cos^2(\theta)  e^{i P \cos^2(\theta)} \bm{\chi_j}
\label{p-field-explicit}
\end{eqnarray}
where $\bm \chi_j$ is a vector of $\langle \phi_n | \chi_j \rangle$. 
In addition, we express
\begin{eqnarray}
    \frac{\partial {\mathcal{O}}_j}{\partial I} = \frac{\partial {\mathcal{O}}_j}
    {\partial \bm h}\frac{\partial \bm h}{\partial I},  
\end{eqnarray}
where $\bm h$ is a vector of eigenenergies, $h_J = J(J+1)/(2I)$. The first term 
can be explicitly written as
\begin{eqnarray}
    \frac{\partial {{\mathcal{O}}}_j}{\partial \bm h} = (i t) \bm u^{\ast\top} \bm{A}_j \bm u - (it) \bm u^\top  
    \bm{A}_j^\top \bm u^\ast,
\label{sj}
\end{eqnarray}
\begin{figure}[ht] 
    \if\flag1\includegraphics{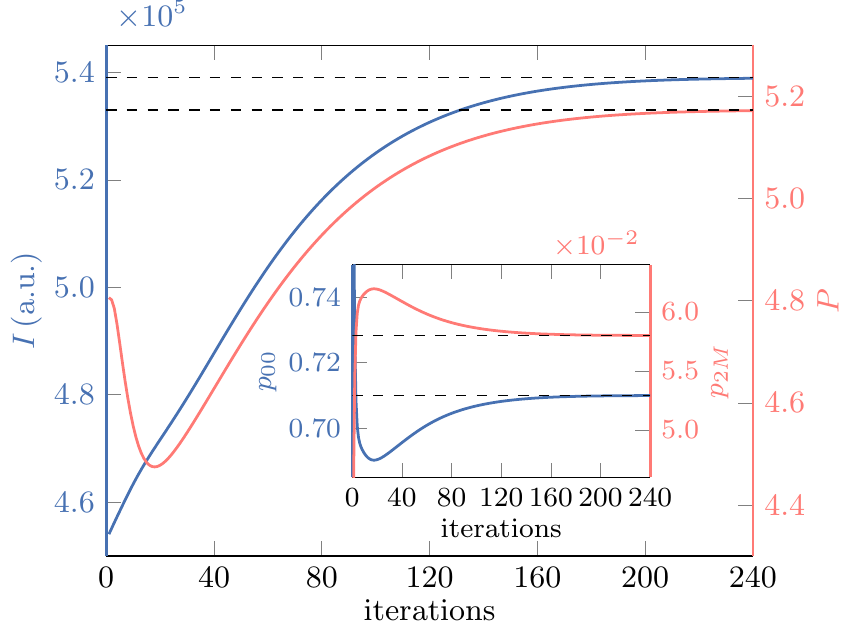}\else\include{fig5}\fi
    \vspace{-3mm}
	\caption{Convergence of the inferred field parameter $P$ and molecular moment
	of inertia $I$, using $\langle \cos^2\theta\rangle(t)$. The inset 
	shows the convergence of the probabilities of $\ket{J=0,M_J=0}$, 
	$\ket{2,-2}$, $\ket{2,-1}$, $\ket{2,0}$, $\ket{2,1}$ and 
	$\ket{2,2}$ in the initial mixed state.
	The dashed lines show the reference values to be inferred. 
	}
\label{P-I-and-prob}
\end{figure}%
where $\bm u$ is a vector of $\exp(ih_Jt)$, $\bm{A}_j = \bm {S}_j^\dagger \bm {{\cal O}} \bm {S}_j$
and $\bm {S}_j$ is a diagonal matrix of $\langle \phi_n | \psi_j \rangle$. The second term is given by 
$\partial h_J/ \partial I = - J(J+1)/(2I^2)$. 
We apply Eqs. (\ref{P-field})--(\ref{C-b}), and (\ref{p-field-explicit})--(\ref{sj}) to evaluate the gradient with respect to $P,\,I,\,p_j$, which are used in projected gradient descent.
We consider an alignment signal 
$\braket{\cos^2(\theta)}(t)$ produced by exciting molecules with $I=539010 \,\mathrm{a.u.}$ 
in a mixture of $J=0$ and $J=2$ states by linearly polarized pulse with $P=5.174$ yielding wave packets 
including rotational states with $J \leq 10$. Figure \ref{P-I-and-prob} illustrates the convergence of $I$, $P$ and 
the probabilities in the initial mixed state to the corresponding inferred values. 
Notice that all $\ket{2,M}$ states are equally populated initially.

\section{Conclusion}
In summary, we have shown that for a time-independent Hamiltonian $\hat{H}$ with 
spectrum parametrized by $\alpha$ and quantum states prepared by $\hat{V}(P)$ over finite time,
it is possible to obtain closed-form expressions for the gradients of the distance
between calculated and observed 
time-dependent signal with respect to $\alpha$, $P$ and the elements of the 
system density matrix. These gradients can be used in projected gradient descent 
to infer $\alpha$, $P$ and the density matrix from dynamical observables, 
such as molecular alignment or orientation. We have shown that this 
approach can be combined with random phase wave function approximation, 
yielding closed-form expressions for gradients that can be used to infer 
population distributions spanning a large number of molecular states from averaged 
time-dependent observables. We have illustrated that the temperature of a molecular 
gas at ambient conditions can be determined optically by probing time-dependence of alignment. This demonstrates 
a new way of probing molecular distributions in applications requiring remote 
sensing and opens new possibilities for solving inverse quantum problems 
with averaged dynamical observables. 

We note that several important questions remain unanswered by this study. 
While our calculations demonstrate that some density matrices can be reconstructed with several simultaneously measured observables,  we believe that 
the minimum number of observables required for the reconstruction of the full density matrix is generally not known. 
Refs. \cite{jamiolkowski1983minimal,jamiolkowski2004stroboscopic} provided a formula for the minimum number of observables based on the degeneracy of the time evolution operator. In particular, it was suggested that a non-degenerate system requires $d$ observables, 
where $d$ is the dimension of the Hilbert space. However, this clearly contradicts our results in Sect. \ref{pl} and the results in Refs. \cite{Hasegawa2008qsr,He2019Direct, karamatskos2019molecular,Ueno2021QS}. 
Further examination is required to investigate what determines the minimum amount of observable information, both in terms of the number of observables and the duration of the time-dependent signals, for the reconstruction of molecular density matrices, including the relationship between the number of observables, symmetries and parameters of the state-preparation fields.  
This question is particularly relevant for extensions of the present approach to include vibrational and electronic degrees of freedom. 

\begin{acknowledgments}
The work of WZ and RVK is supported by NSERC of Canada.
IT and IA thank Israel Science Foundation  (Grant No. 746/15) for supporting this work. 
RVK is grateful to the Weizmann Institute of Science for the kind
hospitality and support during his stay as a Weston Visiting Professor.
IA acknowledges support as the Patricia Elman Bildner
Professorial Chair and thanks the UBC Departments of
Physics \& Astronomy and Chemistry for hospitality extended to him
during his sabbatical stay.
This research was made possible in part by the historic generosity of the Harold Perlman Family.
\end{acknowledgments}
\vspace{5mm}
\section*{Appendix: Derivative of Complex Quadratic Form}\label{sec:App-A}
Consider $f(\bm x)=\bm x^{\dagger}\bm U \bm x$, where $\bm x$ is a complex valued column vector and $\bm U$ is a Hermitian matrix:
\begin{flalign}
    \!f(\bm x) = \sum_{i,j=1}^{n} x_{i} u_{i j} x_{j}=\sum_{i=1}^{n} \Big[ u_{i i} x_{i}^{2}+\sum_{j \neq i} x_{i} u_{i j} x_{j} \Big], & {} \label{eq:App-1}
\end{flalign}
where $x_i$ are the elements of $x$ and $u_{ij}$ are the elements of $U$.
By writing $x_i=a_i+ib_i$, we recast Eq. \eqref{eq:App-1} as
\begin{eqnarray}
  \begin{aligned}
    f(\bm x) =\sum_{i=1}^n &\left[\vphantom{\sum_{i=1}^n} u_{ii}(a_i^2+b_i^2)\right.+  \\
    & \left. \sum_{j \neq i}u_{ij}(a_ia_j+b_ib_j+ia_ib_j-ib_ia_j) \right].\qquad\qquad
  \end{aligned}
\end{eqnarray}
A partial derivative with respect to $a_k$ is
\begin{eqnarray}
\begin{aligned}
  \frac{\partial f}{\partial a_k}  = & \quad 2u_{kk}a_k+\sum_{j \neq k}u_{kj}(a_j+ib_j)+\sum_{i \neq k}u_{ik}(a_i-ib_i)\\
   = & \quad u_{kk}(a_k+ib_k)+u_{kk}(a_k-ib_k)\\
    & \quad +\sum_{j \neq k}u_{kj}(a_j+ib_j)+\sum_{i \neq k}u_{ik}(a_i-ib_i) \\
   =& \quad \sum_{j=1}^n u_{kj}(a_j+ib_j) + \sum_{i=1}^n u_{ik}(a_i-ib_i),
\end{aligned}
\end{eqnarray}
which yields
\begin{eqnarray}
\nabla_{\bm a} f(\bm x) = \left[ \frac{\partial f}{\partial a_k}\right] = \bm U \bm x+\bm U^\top \bm x^*.
\end{eqnarray}
Similarly, 
\begin{flalign}
  \frac{\partial f}{\partial b_k} = -i\sum_{j=1}^n u_{kj}(a_j+ib_j) + i\sum_{i=1}^n u_{ik}(a_j-ib_j), & {} 
\end{flalign}
yielding
\begin{eqnarray}
\nabla_{\bm b} f(\bm x) = \left[ \frac{\partial f}{\partial b_k}\right] = i\bm U^\top \bm x^*-i \bm U \bm x.
\end{eqnarray}

\clearpage
\end{document}